# Excitation and Imaging of Resonant Optical Modes of Au Triangular Nano-Antennas Using Cathodoluminescence Spectroscopy

Anil Kumar[1], Kin-Hung Fung[2], James C. Mabon[3], Edmond Chow[4], and Nicholas X. Fang[2]*

[1]*Department of Electrical and Computer Engineering, University of Illinois at Urbana-Champaign, 1406 W. Green Street, Urbana, IL 61801 USA*

[2]*Department of Mechanical Science and Engineering, University of Illinois at Urbana-Champaign, 1206 W. Green Street, Urbana, IL 61801 USA*

[3]*Fredrick Seitz Materials Research Laboratory, University of Illinois at Urbana-Champaign, 104 S. Goodwin Avenue, Urbana, IL 61801 USA*

[4]*Micro and Nanotechnology Laboratory, University of Illinois at Urbana-Champaign, 208 N. Wright Street, Urbana, IL 61801 USA*

*Electronic mail: nicfang@illinois.edu

Cathodoluminescence (CL) imaging spectroscopy is an important technique to understand resonant behavior of optical nanoantennas. We report high-resolution CL spectroscopy of triangular gold nanoantennas designed with near-vacuum effective index and very small metal-substrate interface. This design helped in addressing issues related to background luminescence and shifting of dipole modes beyond visible spectrum. Spatial and spectral investigations of various plasmonic modes are reported. Out-of-plane dipole modes excited with vertically illuminated electron beam showed high-contrast tip illumination in panchromatic imaging. By tilting the nanostructures during fabrication, in-plane dipole modes of antennas were excited. Finite-difference time-domain simulations for electron and optical excitations of different modes showed excellent agreement with experimental results. Our approach of efficiently exciting antenna modes by using low index substrates is confirmed both with experiments and numerical simulations. This should provide further insights into better understanding of optical antennas for various applications.

## I. INTRODUCTION

Recent advances in resonant sub-wavelength optical antennas[1-4], as an optical counterpart of microwave antennas, has offered researchers a continuum of electromagnetic spectrum to design, analyze and predict new phenomena that were previously unknown[1, 4-5]. Their applications in areas with pressing needs, e.g., in sensing[4], imaging[6,7], energy harvesting, and disease cure and prevention have brought revolutionary improvements. However, understanding of the physics behind the optical interaction with metal nanostructures is still in the early stage. By patterning metal films in unique shapes and sizes, highly localized optical modes can be engineered. Exciting and observing these modes require state-of-the-art techniques, e.g., near-field scanning optical microscopy (NSOM), electron energy loss spectroscopy (EELS), and cathodoluminescence (CL) spectroscopy. Among these, CL is advantageous because of its high resolution, ease-of-excitation, and less stringent demands for sample preparation[8]. Because the impinging electrons offer all the possible wave vectors, momentum matching can be easily achieved which can be an important constraint for optical excitation of nanostructures.

CL has previously been used to study optical modes of nanoantennas[8-11]. An important concern for CL has been the strong luminescence from the substrate that can over-shadow any possible antenna modes. Most commonly used materials in optics and device processing show luminescence under electron excitation making the nature and purity of the substrate critical. As an example, float glass shows eight peaks in the visible spectrum[12] making it almost impossible to distinguish any antenna peaks. To avoid the problem of substrate signal, high purity materials, e.g., Si, InP, or other single crystals that do not show any peaks in visible region can be used; however, their high index shifts the dipole peaks beyond the detection range of the setup (Photomultiplier tube, 250-900 nm). Since resonance wavelength of an antenna is proportional to substrate index and particle size, the high index places a huge constraint on fabrication. The characteristic length of optical antennas is of the order of hundred nanometers or less—this constraint in terms of fabrication may not be easy to overcome and any reduction in index design can have significant effect in studies of optical nano-antennas. Here we take a new approach to design, excite and observe optical modes of triangular Au nanoantennas using CL spectroscopy. Both out-of-plane as well as in-plane dipole modes—which have previously been very difficult to excite on flat nanostructures using CL spectroscopy—are reported. This approach based on nearly-free standing nanostructures is not limited to triangles but can be used for any structure and we show how it can help overcome some of the material and design constraints CL places on sample preparation and e-beam excitation.

## II. SAMPLE DESIGN

To obtain a low index substrate with minimal background luminescence, we fabricated gold nanostructures on 100 nm thick PECVD-deposited $SiO_2$ film on Si using electron-beam lithography. $SiO_2$ also shows two peaks in this region[13]; however, the high energy peak lies outside the dipole peaks of our designs (and was observed to be relatively weaker for PECVD deposited oxide), while the low energy peak is red-shifted from the antenna peaks. The $SiO_2$ outside the nanostructures was removed using a standard reactive ion etching (RIE) process (PlasmLab Freon-RIE with 60 sccm flow rate of $CHF_3$, at a pressure of 35 mTorr and 90W power). A 50:1 $H_2O$:HF wet etching solution was used for further undercutting of the oxide underneath the nanostructures to provide an effective index close to vacuum. The etching was carefully controlled until the structures were about to fall-off from the oxide triangle. Fig. 1 shows an SEM image of a 50 nm thick Au triangle dimer showing the undercut region. As can be seen from the right triangle, the $SiO_2$ structure has shrunk and is smaller than the Au triangle due to selective wet etching. Beside this shrinking, the solution appears to etch Au-$SiO_2$ interface faster than bulk oxide, as seen from the extra etching of the exposed interface of the left triangle. This results into faster undercutting of the oxide near Au nanostructure which results into a very small contact area. An overall effect of this etching behavior on the antenna is that most of it is surrounded by vacuum, except the small bottom contact with oxide. This translates into an effective index that is close to vacuum instead of an average value observed for a flat substrate.

To predict the effect of substrate on resonant wavelength, we carried out FDTD simulations for various substrates using a plane wave source incident normal to the antenna surface. Resonance peaks as a function of substrate index for triangles with tip-to-base height of 120 nm and 150 nm are shown in Fig. 2. The narrow window of substrate index observed—for which the antenna peaks are in the visible range—highlights the importance of substrate index in optical antennas. Modification of substrate index, $n$, by lifting up the Au nanostructure is also considered. For a 150 nm triangle sitting on a flat $SiO_2$ substrate ($n$=1.5) the simulated peak position is at 700 nm, which shifts to 680 nm when it is lifted up by a 100 nm thick triangle of same size. This configuration reduces the effective index of the substrate resulting in a blue shift which will be further shifted by the undercutting of oxide as shown in Fig. 1.

To understand the role of the low-index configuration in CL measurements, we simulated electron beam excitation of the antennas using FDTD approach which is reported earlier[11]. In summary, the electron-beam was modeled as a series of dipoles with a temporal phase delay based on electron beam velocity. Si was considered as a dispersive medium and dispersion data for Au were taken from the literature[14]. Based on this approach, Fig. 3 compares a 120 nm triangle sitting directly on flat Si surface with another triangle that is lifted up by a 100 nm thick $SiO_2$ triangle of same size. For simplicity we have not included any undercutting in the simulations. The blue-shifted low intensity peak for the Au on Si substrate case indicates this peak is possibly from one of the higher order modes while the dipole mode has shifted beyond the detection range. The high intensity peak for the Au triangle sitting on oxide triangle falls within the range of the CL detector and is about 15-times stronger than the flat Si substrate. This enhancement can be understood considering that when the antenna is sitting close to a high index substrate, most of the field will be radiated into the high-index substrate; the lifting-up and undercutting allows bringing more of this radiation away from the substrate. This approach can be generalized, however, it should be kept in mind that near-field will be modified by the small $SiO_2$ region. Careful analysis of systematic undercutting and its effect on the modes of various nanostructures are currently being pursued.

## III. EXPERIMENTAL RESULTS AND DISCUSSION

CL measurements were carried out on a commercial setup (MonoCL from Gatan) with a parabolic mirror for photon collection mounted onto an SEM (JEOL JSM-7000F) and the spectra were taken using a photomultiplier tube (PMT) with a range of 250 nm – 900 nm. A schematic of the setup is shown in Fig. 1 (a). Fig. 4 shows a spectrum on Au film (black curve) taken at excitation voltage of 30 kV and beam current of 100 nA. The surface plasmon peak around 520 nm is clearly observed as reported earlier[15]. For comparison, spectrum taken over the substrate with $SiO_2$ residue shows two prominent peaks (gray), coming from the residue oxide on Si[13]. Furthermore, measurements performed on a 50 nm thick Au triangle (red curve) with tip-to-edge length, L = 150 nm show a peak at 575 nm. However, this peak does not match with the simulated in-plane modes shown earlier in Fig. 2. Since excitation of higher modes is much more difficult, we believe this peak is due to out-of-plane mode. A similar observation has earlier been associated with an out-of-plane dipole mode for Ag triangles[11]. For comparison, Fig. 4 also shows spectrum taken for a triangle on Si substrate (blue). A weak peak at higher frequency, as suggested in the simulation is observed.

Panchromatic CL image taken by raster scanning of the electron beam at 30 kV and 85 nA is shown in Fig. 4 (inset) and shows high-contrast tip modes. This contrast is much higher for the low index configuration designed here compared to Si substrates which can be confirmed from the spectra. This spatial mode has earlier been associated with an out-of-plane dipole mode for Ag triangles[11]. The panchromatic images can be used to find the resolution limit of CL imaging technique. From images of closely spaced triangle dimers the spatial resolution was found to be approximately 22 nm (FWHM), which is comparable to previously reported value[11,16].

For higher photon collection efficiency the sample needs to be placed very close to mirror. It has earlier been shown that due to the normal incidence of the electron beam onto the particle in such a setting, the resulting out-of-plane dipole will have an electric field profile along the beam direction[11] and predominantly out-of-plane modes are excited. This is less interesting because the resonant behavior is predominantly determined by metal thickness instead of nanostructure geometry. One way to overcome this issue is to orient the structure such that the electron beam is parallel to the plane of triangle. In our current design, further undercutting of oxide allows tilting of the nanostructures relative to beam direction. Fig. 5 shows SEM images of such a tilted triangle. Fig. 5 (a) shows a secondary electron beam image where both the $SiO_2$ and tilted Au nanostructures can be seen, while the backscattered electron image in Fig. 5 (b) resolves the Au structure. Because of the oblique incidence of the beam, panchromatic images taken on this triangle can still show the spatial modes of the antenna, as seen in Fig. 5 (c).

Next, we explore the possibility of exciting in-plane modes of the triangle antennas using a beam parallel to the surface. Due to rotational symmetry, the in-plane mode of triangular nanoantenna is independent of triangle rotation[17]. Therefore, the in-plane rotation of the triangle is not critical, as long as the effect of substrate on triangles sitting on tip or edge is comparable. Spectrum taken close to a 120 nm Au triangle with near-vertical tilt shows a peak at about 600 nm (Fig. 6). This in-plane peak position shows very good match with the extinction peak predicted from FDTD simulations for an Au triangle in vacuum (Fig. 2).This is further confirmed by FDTD simulation using an electron beam excitation (red curve, Fig. 6). The additional peaks appear due to the $Si/SiO_2$ background as can be seen from the gray curve taken over the $SiO_2$ nanostructure. The relatively wider peak of the CL spectra is possibly due to broadening of the excited mode when the triangle is not fully oriented in vertical direction and will be affected by fabrication imperfections. Additionally, the simulated particle has a vacuum surrounding resulting into narrower peaks, while the experimental peak width will increase due to Si substrate at the bottom.

For a fully tilted triangle, this approach does not allow imaging of the in-plane mode. Therefore, to understand the nature of excitation using an electron beam parallel to the surface of the structure, we simulated electric field pattern of a triangular antenna sitting in air with electron beam parallel to the surface, as shown in the schematic of Fig. 6 (b). The electric field 2 nm below the lower interface is plotted in Fig. 6 (c) showing a resonance behavior of the nanoantenna similar to plane-wave excitation, confirming our initial assumption as well as experimental observation that an electron beam parallel to the surface should be able to excite in-plane modes of the antenna.

## IV. CONCLUSION

Au triangles were fabricated on substrates with low effective index and minimal background luminescence. CL spectroscopy carried out on these triangular nanoantennas allowed spatial and spectral mapping of various modes. High contrast tip modes in panchromatic images were observed which were associated with an out-of-plane dipole mode. In-plane modes were excited by orienting the flat surface of triangle parallel to the electron beam which was achieved by extra undercutting of the oxide triangles below the Au antenna. Excellent matching with the FDTD simulations using electron-beam as well as plane wave excitations was shown. This design approach was found to be more efficient for excitation of plasmonic modes of gold triangles and should allow imaging of various modes associated with other optical nanoantennas.

## ACKNOWLEDGEMENT

The authors would like to thank Dr. Jun Xu and Dr. Pratik Chaturvedi for helpful discussions. This work was carried out in part in the Frederick Seitz Materials Research Laboratory Central Facilities, University of Illinois, which are partially supported by the U.S. Department of Energy under grants DE-FG02-07ER46453 and DE-FG02-07ER46471.

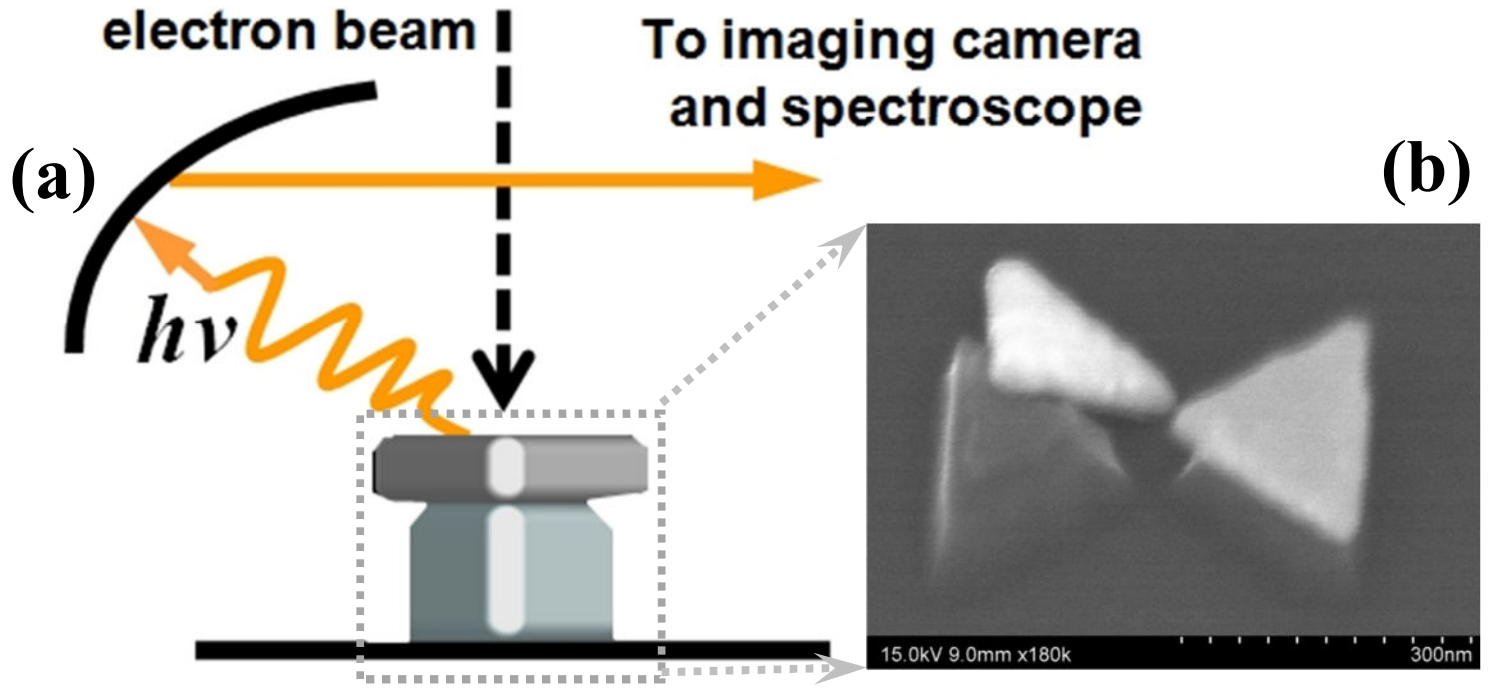

Fig. 1. (a) Schematic of the cathodoluminescence setup. (b) SEM image of an Au triangle-dimer on 100 nm thick $SiO_2$ deposited on Si. After e-beam lithography to fabricate the 50 nm thick Au structures, $SiO_2$ outside the nanostructures was removed using Freon-RIE and further undercutting was achieved using wet etching.

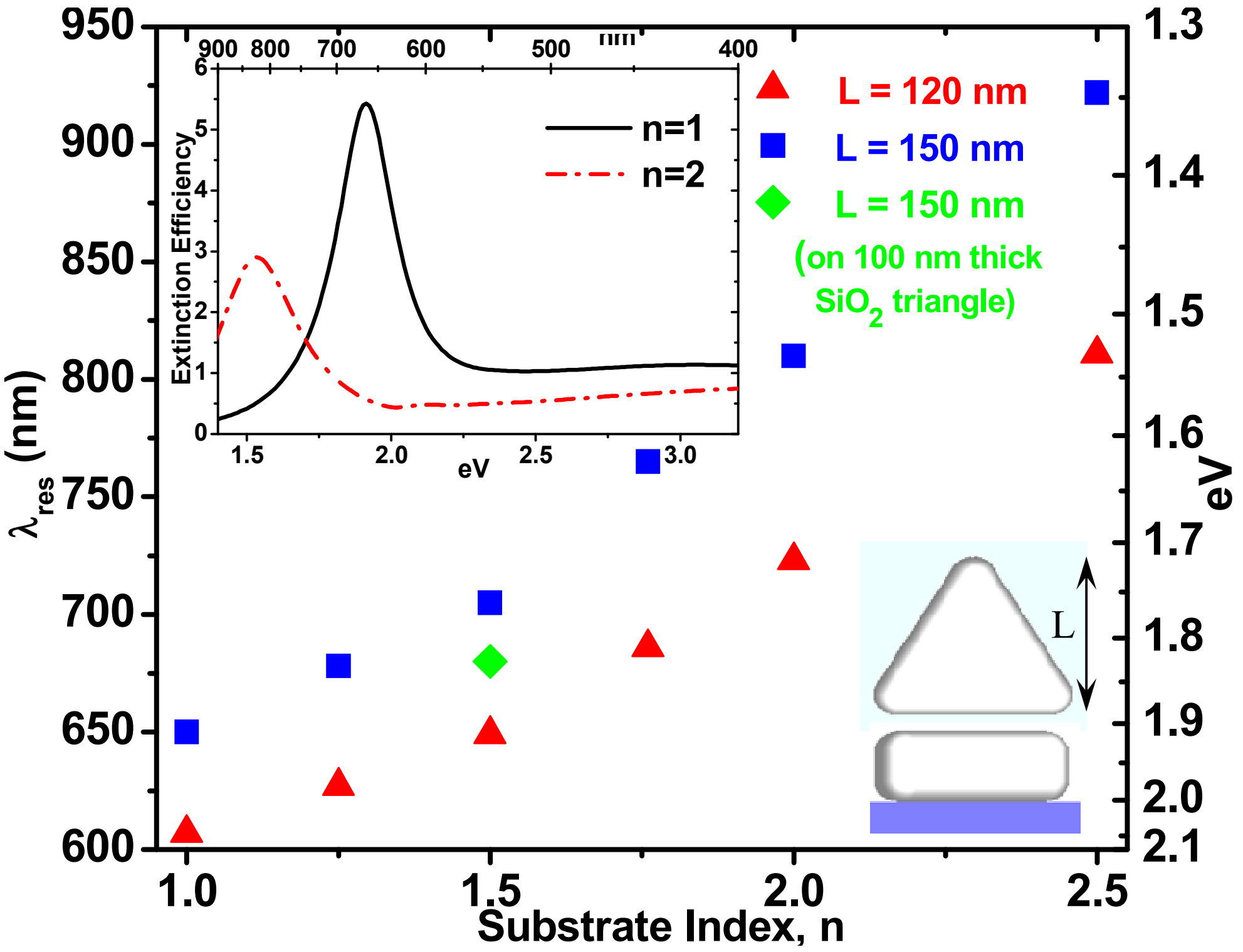

Fig. 2. Simulated in-plane dipole resonance peaks versus substrate index of Au triangular nano-antennas using plane-wave incidence. (Inset) Representative extinction spectra for a triangle (L = 150 nm) for two different substrate indices. A 15 nm curvature was used for the triangle edges and tips.

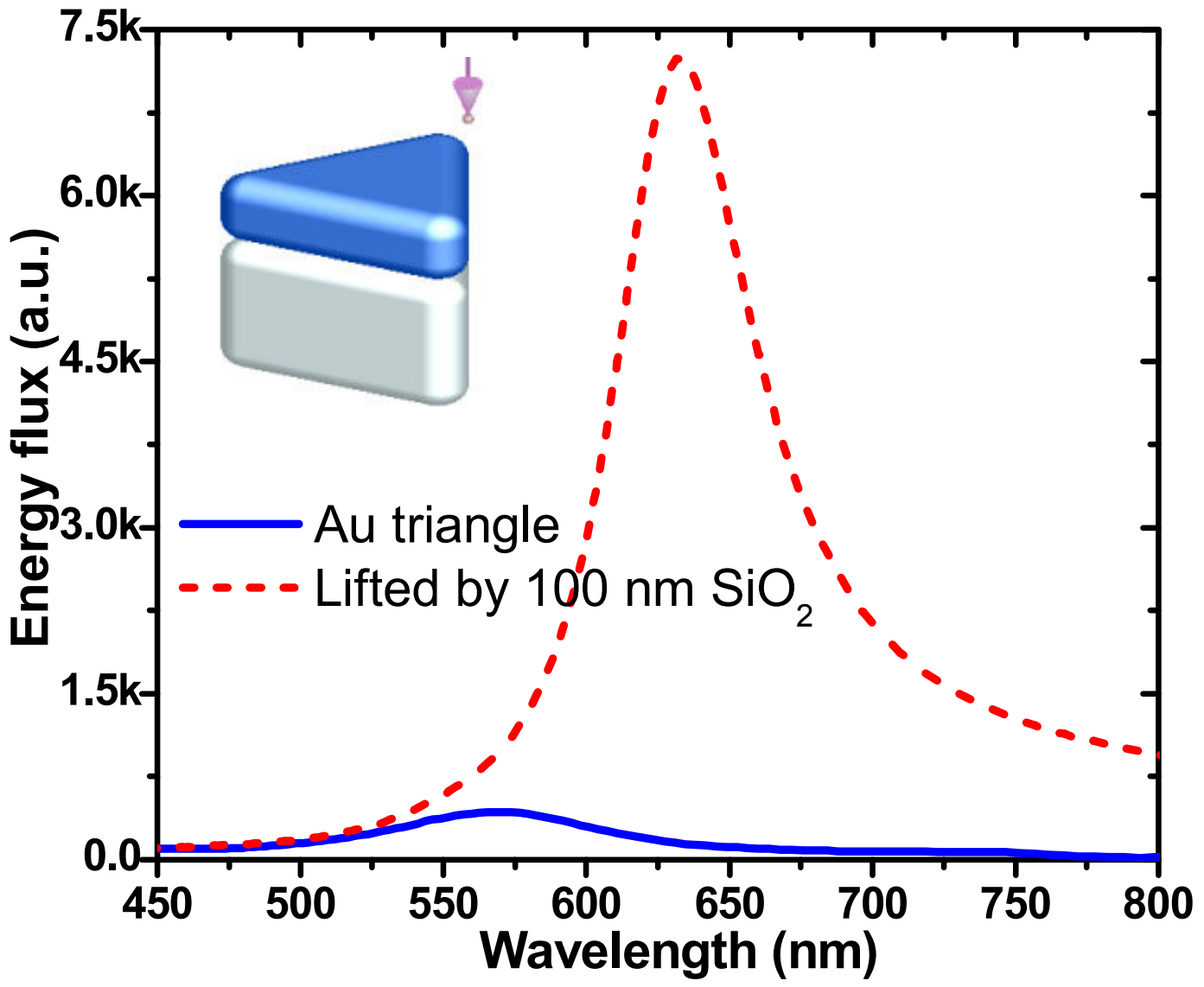

Fig. 3. FDTD simulation comparing signal from an electron beam excitation when the beam is focused near the tip of a 120 nm triangle with and without 100 nm thick $SiO_2$ triangle on Si substrate.

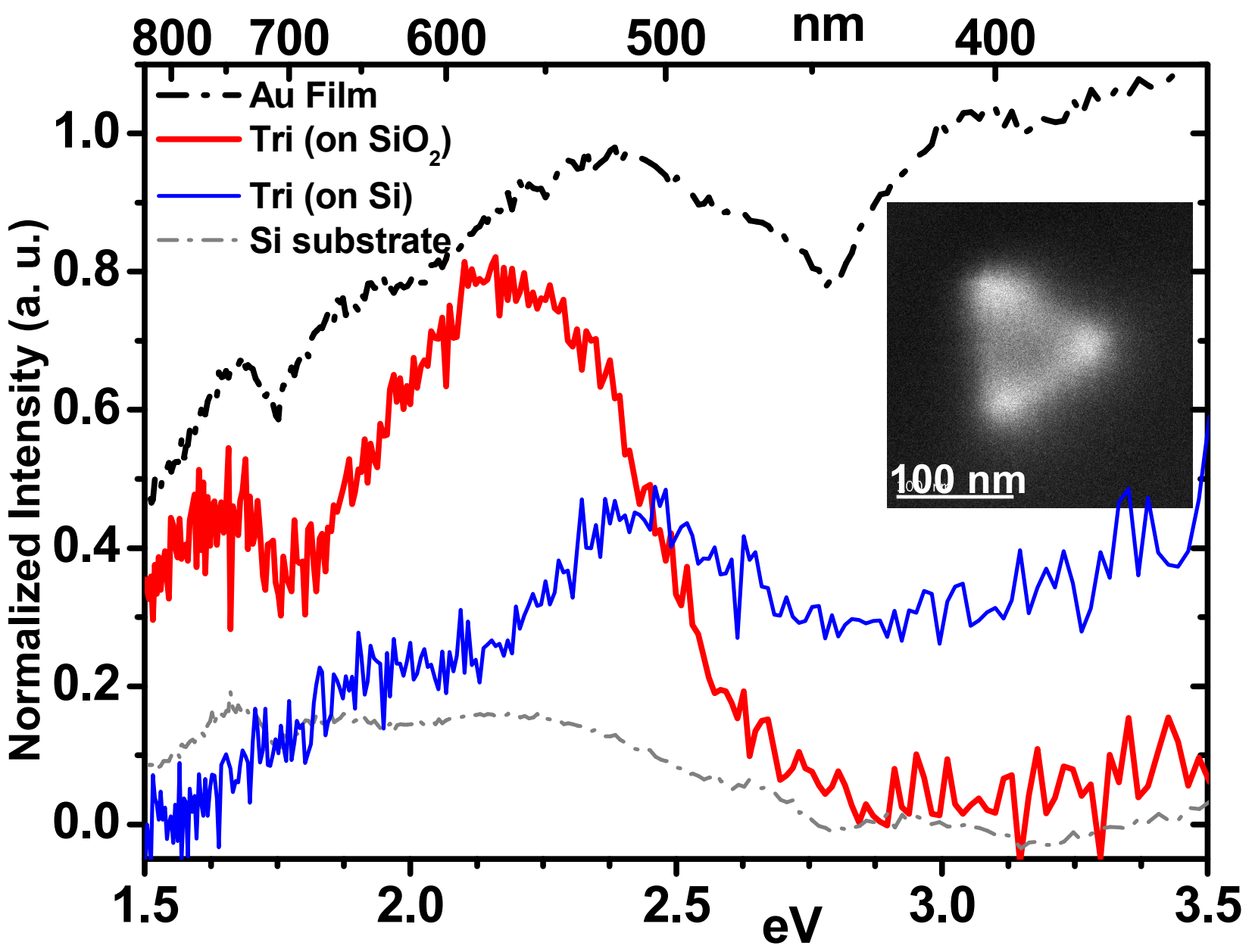

Fig. 4. CL spectrum for Au triangle (red) taken by raster scanning of the electron beam and shown after grating correction. Spectrum for Au film (black) with surface plasmon peak at 520 nm[15] is shown for comparison and a background curve (gray) shows two prominent peaks related to oxide[13]. The curves are shifted relative to the normalized red curve for comparison. Inset shows a panchromatic image of Au triangle after oxide removal using Freon-RIE and wet etching.

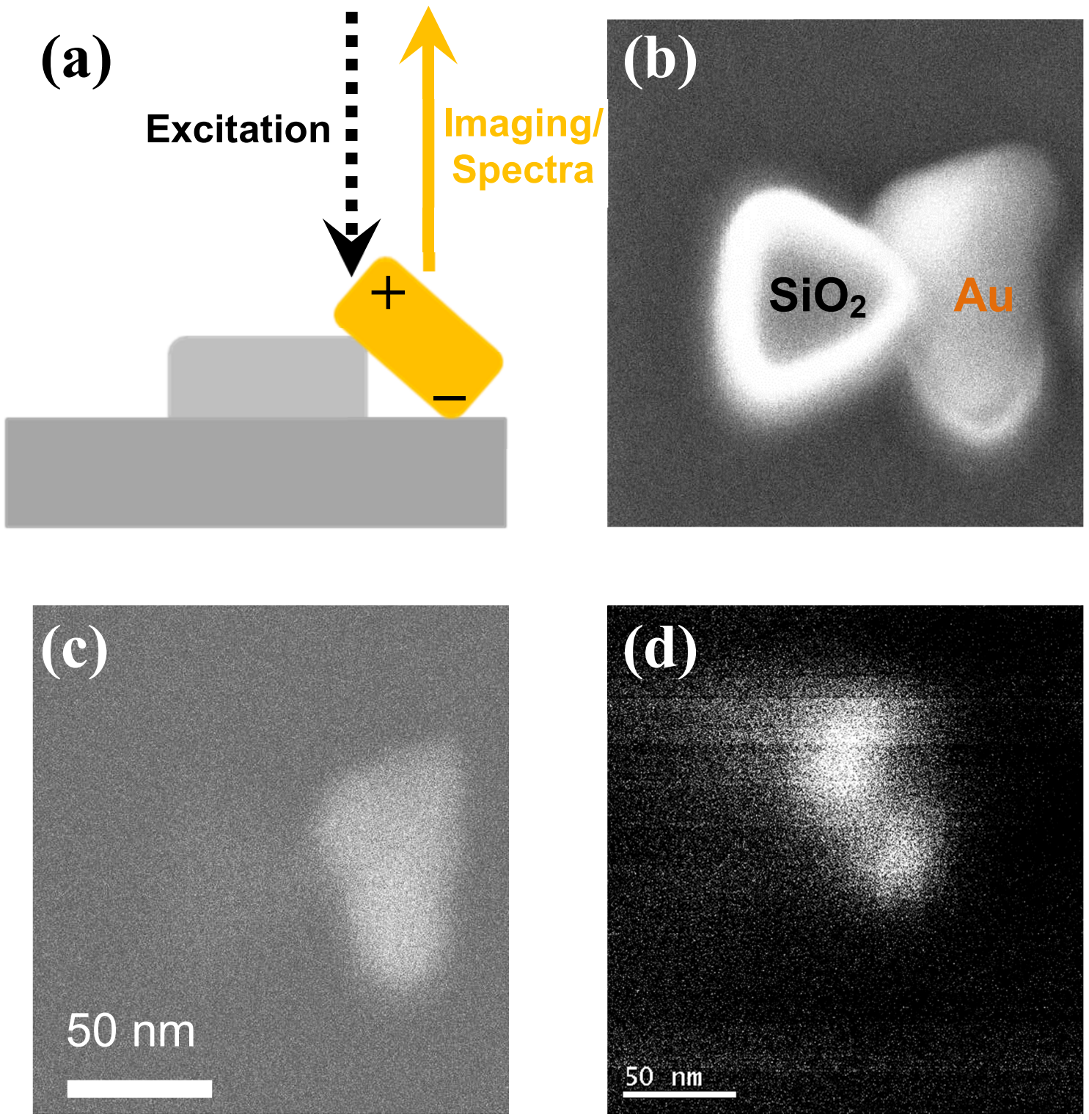

Fig. 5. Imaging plasmonic modes of tilted triangles: (a) shows schematic of the approach taken; (b) a secondary electron image resolving both $SiO_2$ and the fallen Au triangle, while (c) shows the backscattered electron image which preferentially shows heavy metals (Au triangle). (d) a panchromatic CL image of this triangle shows the tip modes allowing both spectral as well as spatial mode imaging.

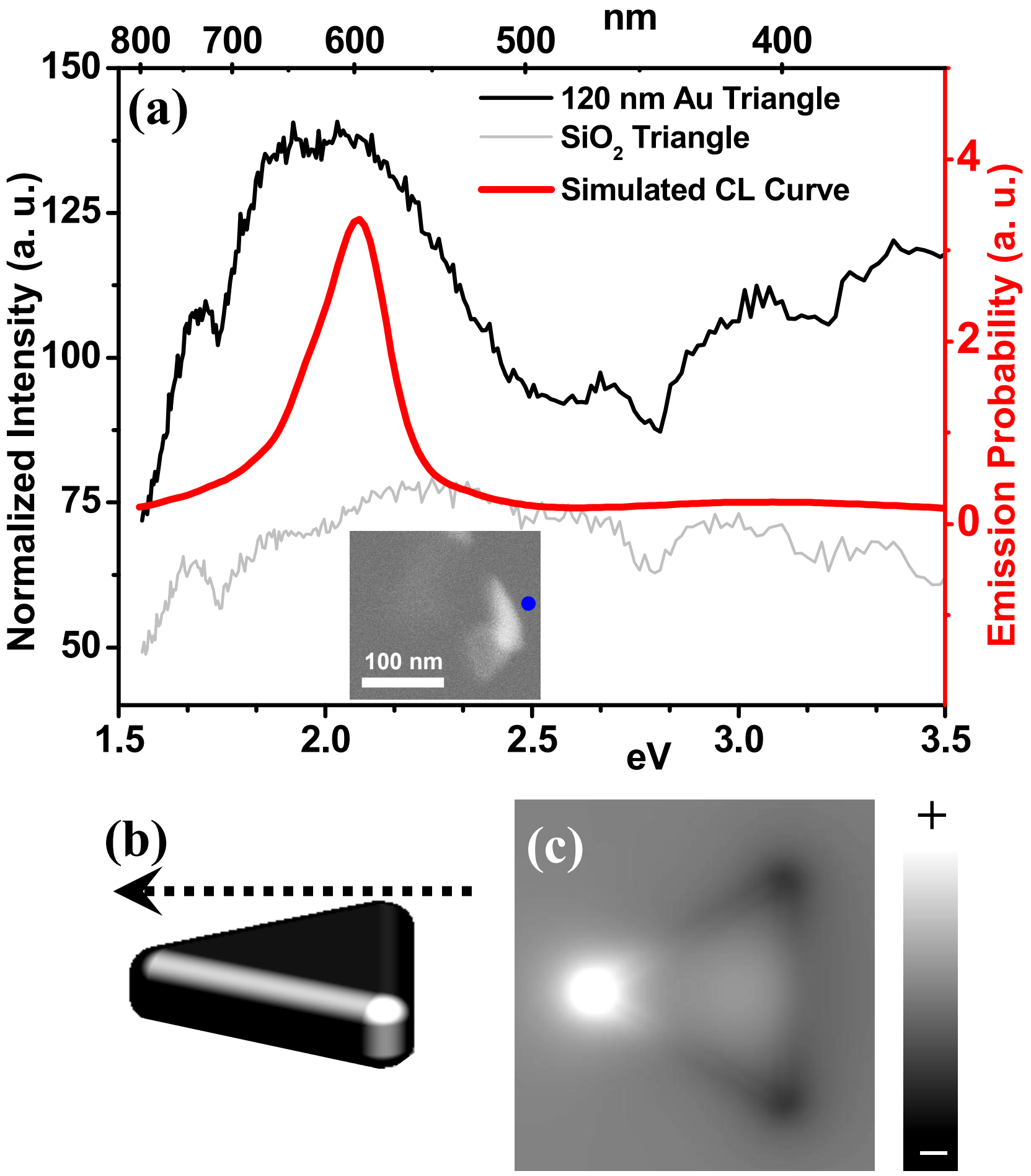

Fig. 6. CL measurements for in-plane mode excitation on a tilted triangle (black) shows excellent agreement with an FDTD simulated curve for electron-beam excited triangle in vacuum (red). Inset shows a backscattered SEM image of 90° tilted triangle with electron-beam position (blue dot). (b) shows schematic of simulation conditions with electron beam parallel to the surface of the triangle (shown as an arrow). (c) electric field profile of the excited mode 2 nm away from the surface (opposite to the electron beam).